# Design and Fabrication of Micromachined Resonators

Ritesh Ray Chaudhuri[a], Joydeep Basu[b], Tarun Kanti Bhattacharyya[*b]
[a] Advanced Technology Development Centre, Indian Institute of Technology, Kharagpur, India
[b] Dept. of Electronics and Electrical Communication Engg, Indian Institute of Technology, Kharagpur, India

## ABSTRACT

Microelectromechanical system (MEMS) based on-chip resonators offer great potential for sensing and high frequency signal processing applications due to their exceptional features like small size, large frequency-quality factor product, integrability with CMOS ICs, low power consumption etc. This work is mainly aimed at the design, modeling, simulation, and fabrication of micromachined polysilicon disk resonators exhibiting radial-contour mode vibrations. A few other bulk mode modified resonator geometries are also being explored. The resonator structures have been designed and simulated in CoventorWare finite-element platform and fabricated by the PolyMUMPs surface micromachining process.

## 1. INTRODUCTION

One of the primary focuses for research and development in the area of RF wireless communications electronics has been device miniaturization. Frequency-reference circuits are ubiquitous in modern integrated electronic systems. Microelectromechanical system (MEMS) based resonators provide a feasible alternative to the present-day well-established quartz crystal technology that is riddled with major drawbacks like relatively large size, high cost, and poor compatibility with integrated electronic circuits. MEMS based resonators transduce electrical signals into extremely low-loss mechanical vibrations and vice versa, and offer great potential for sensing and high frequency signal processing applications due to their exceptional features like small size, large frequency-quality factor product, integrability with CMOS integrated circuits, low power consumption, adequate temperature and ageing stability, low cost batch fabrication etc. Both capacitive [1, 2] and piezoelectrically transduced [3] versions of such resonators have been demonstrated.

Due to the orientation-independent Young's modulus and higher mechanical quality factor of polysilicon than single-crystal silicon, the former is an attractive structural material for the fabrication of microresonators. Micromechanical resonators like beams, square-plates, circular disks etc. can vibrate in either bulk-mode or flexural-mode, with bulk-modes being preferred for high frequency generation due to their larger structural stiffness [4]. Bulk-mode vibration of circular disks is again possible in two distinct modes as illustrated in Fig. 1: (a) Radial-contour (or, breathing) mode where the shape of the disk expands and contracts equally in all the lateral surface and (b) Elliptical (or, wine-glass) mode where the disk expands along one axis and contracts in the orthogonal axis forming two alternate and perpendicular ellipses per cycle of vibration with four nodal points at the perimeter. Radial-contour modes provide higher effective stiffness and hence, are preferred [1].

---

[*b] tkb@ece.iitkgp.ernet.in, +91- 322 - 283554

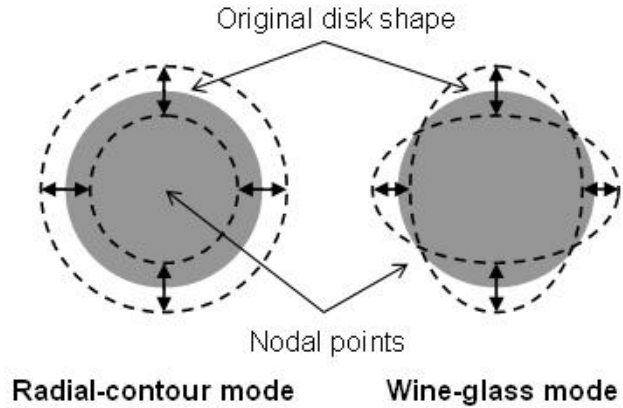

Figure 1. The two modes of bulk acoustic vibration of a disk (in the first or, fundamental mode). The dotted curves show the fully expanded and contracted mode shapes

This paper is focused on radial-contour bulk-mode circular-disk shaped MEMS resonators which depend on capacitive actuation and readout technique. Along with this, a few modified resonator geometries have also been explored that can provide us with improved performance. The resonator structures have been designed and simulated in CoventorWare finite-element platform and fabricated by the PolyMUMPs surface micromachining process.

## 2. THEORY

The resonator in this work consists of a circular-disk made of polysilicon suspended by a cylindrical stem at its center which is a nodal-point for the radial-contour vibrations. Along its perimeter, the disk is surrounded by lateral capacitive-gap input and output electrodes (also made of polysilicon), which act as electromechanical transducers. Its schematic diagram is shown in Fig. 2.

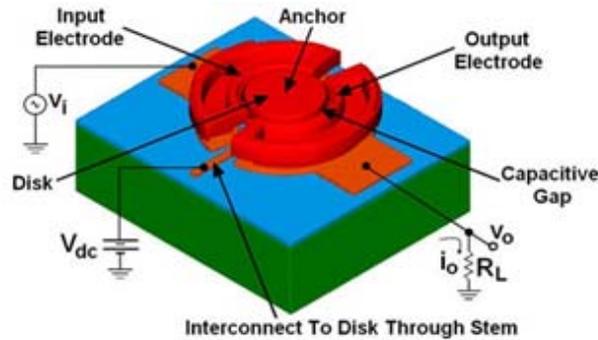

Figure 2. Schematic diagram of the disk resonator

The resonator is excited into forced-vibrations by applying a sinusoidal voltage $v_i$ at the disk's resonance frequency $f_0$ to the input electrode and a dc-bias voltage $V_{dc}$ to the disk through its stem, hence producing a time-varying radial electrostatic force on the disk [1]. Due to the symmetrical expansion and contraction of the disk around its perimeter, there occurs a change in the disk-to-output electrode capacitance (which is dc-biased by $V_{dc}$) with time, hence producing a sinusoidal motional current $i_o$ with frequency $f_0$ in the load connected at the output port. Thus, an electrical signal at the input port gets transduced into a mechanical signal (force on the resonator) which is filtered by the high-Q mechanical response of the disk giving sinusoidal displacement of it at $f_0$. This mechanical response is translated back to the electrical domain by the output transducer.

### 2.1 Resonance Frequency

The resonance frequency of the disk resonator can be derived in terms of Bessel functions [1, 5]:

$$\frac{J_0\left(\frac{\varsigma}{\xi}\right)}{J_1\left(\frac{\varsigma}{\xi}\right)} = (1-\sigma) \tag{1}$$

$$\varsigma = \omega_0 R\sqrt{\frac{\rho(2+2\sigma)}{E}}, \quad \xi = \sqrt{\frac{2}{1-\sigma}} \tag{2}$$

where, $J_\alpha$ is Bessel function of the first kind of order $\alpha$, $\omega_0$ is the angular resonance frequency, R is the radius of the disk, and E, $\rho$ and $\sigma$ are the Young's modulus of elasticity, mass density and Poisson's ratio of the material of the disk respectively. For polysilicon, E = 160 GPa, $\rho$ = 2,300 kg/m$^3$ and $\sigma$ = 0.22. Simplification of (1) and (2) can yield the following expression for the resonant frequency for the i$^{th}$ breathing mode:

$$\omega_0 = \frac{\lambda_i}{R}\sqrt{\frac{E}{\rho(1-\sigma^2)}} \tag{3}$$

Here, $\lambda_i$ is the frequency parameter for the i$^{th}$ mode ($\lambda_1$ = 1.99, $\lambda_2$ = 5.37, $\lambda_3$ = 8.42, $\lambda_4$ = 11.52) [5]. A plot showing the nature of variation of the resonance frequencies of the first four radial-contour modes with disk-radius is given in Fig. 3.

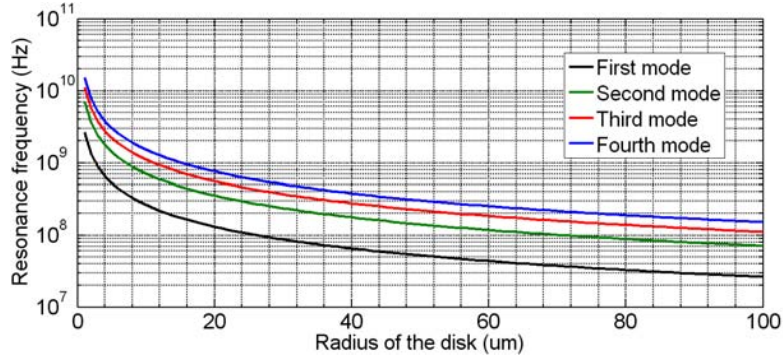

Fig. 3. Variation of radial-contour mode resonance frequency with radius of the resonator disk

### 2.2 Mechanical Model

The resonator can be represented by a mechanical lumped-element model of a mass-spring-damper system having a single degree-of-freedom (Fig. 4(a)) utilizing the total kinetic energy $E_k$ of the vibrating disk which is stated below:

$$E_k = \frac{1}{2}\rho t \int_0^R \int_0^{2\pi} r v^2(r)\, dr\, d\theta \tag{4}$$

Here, v(r,$\theta$) is the velocity at any point (r,$\theta$) on the resonator disk. Thus, expression for the effective mass of the vibrating disk is:

$$m_{eff} = \frac{2E_k}{v^2(R)} = \frac{2\pi\rho t \int_0^R r J_1(hr)^2\, dr}{J_1(hR)^2} = \pi\rho t R^2\left[1 - \frac{J_0(hR)J_2(hR)}{J_1(hR)^2}\right] \tag{5}$$

with

$$h = \omega_0 \sqrt{\frac{\rho}{\left(\frac{E}{1+\sigma}\right) + \left(\frac{E\sigma}{1-\sigma^2}\right)}} = \frac{\lambda_i}{R} \tag{6}$$

The effective spring-stiffness of the resonator $k_{eff}$ (or, $k_r$) is related to the resonance frequency by the relation $\omega_0 = \sqrt{k_{eff}/m_{eff}}$. Also, the damping factor $b_{eff}$ that takes into account the energy-losses of the system is given by:

$$b_{eff} = \frac{\omega_0 m_{eff}}{Q} = \frac{\sqrt{k_{eff} m_{eff}}}{Q} \tag{7}$$

## 2.3 Electrical Model

We can also have a lumped-element series-resonant RLC electrical model for the resonator [1]. The elements of the electrical circuit can be related to those of the mechanical model by using the 'force-voltage analogy' as follows:

$$l_e = m_{eff},\ r_e = b_{eff},\ c_e = (1/k_{eff}) \tag{8}$$

The electromechanical transduction at the two ports ($k = 1, 2$) can be modeled by transformers with turns-ratio $n_k$ (called the electromechanical-coupling coefficient):

$$n_k = V_{dc} \frac{\delta C_k}{\delta r} = V_{dc} \frac{\delta}{\delta r}\left(\frac{\varepsilon A_k}{d_0 - r}\right) \approx V_{dc}\left(\frac{\varepsilon A_k}{d_0^2}\right) (r \ll d_0) \tag{9}$$

where $C_k$ denotes the capacitance between the disk and the $k^{th}$ electrode, $A_k$ is the coupling area given by ($\varphi_k R t$), $\varphi_k$ is the angular overlap of the electrode with the disk, $t$ is the disk thickness, $d_0$ is the static transducer gap, $\varepsilon$ is the dielectric constant of the material between the disk and electrodes, and $r$ is the lateral displacement. By reflecting the motional elements through the transformers at the I/O-ports, we can arrive at a simpler version of the equivalent circuit of the resonator as depicted in Fig. 4(b). The corresponding element values are as follows (assuming symmetrical electrodes i.e., $n_1 = n_2 = n$):

$$L_e = (l_e/n^2),\ R_e = (r_e/n^2),\ C_e = n^2 c_e \tag{10}$$

Also, $C_{0k}$ represents the capacitance between each electrode and the ac-ground.

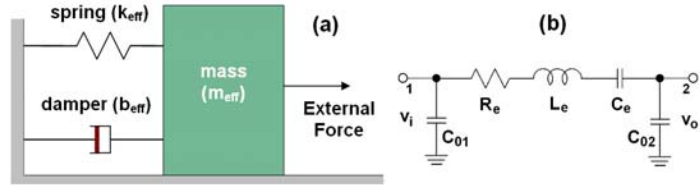

Fig. 4. Equivalent mechanical (a) and electrical (b) model of a MEMS resonator

## 3. SIMULATION RESULTS

Modeling of the micromechanical resonators has been done using Finite-Element Method (FEM), which is a computational method that can simulate the physical and electrical responses of the system almost accurately. The simulation is performed using the FEM software tool *CoventorWare*.

The first three modes for radial-contour vibrations of the disk are illustrated in Fig. 5; having one, two and three nodes respectively. For all these modal displacement simulations, we have taken the bottom surface of the stem to be fixed. Here, the vibration of the disk is entirely within the plane of the device, with ideally no out-of-plane motion. For modes of higher order, additional nodal circumferences gradually get added at which the resonator is also stationary, and the phase of vibration reverses. Analytical solutions have been compared with those obtained from FEM-based simulations for 2 μm thick polysilicon disks of varying radii, and for the first two breathing modes. This is illustrated in Fig. 6.

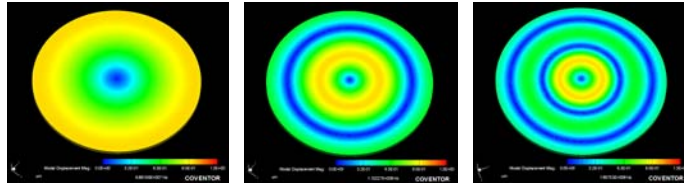

Figure 5. Contour plots for a disk vibrating in the radial-contour mode up to the third order, with the color indicating the modal displacement magnitude.

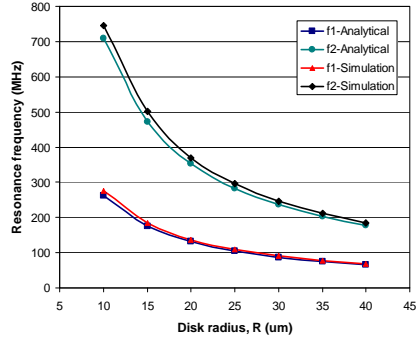

Figure 6. Resonance frequencies for the first (f1) and second (f2) radial-contour modes versus the radius of a polysilicon disk obtained analytically and from simulation.

The structural response of the disk resonator subjected to a harmonic excitation is examined in harmonic analysis. The corresponding simulation is performed by exciting the microresonator disk with an external harmonic load applied to its lateral surface. The frequency of the force is swept to find the resonance frequency with maximum displacement. Fig. 7 shows the generalized displacement plot obtained from harmonic simulation of a polysilicon disk of 40 μm radius and 2 μm thickness, with a 100 kPa load applied to the side-surface of the disk. As the fundamental radial-contour mode resonance frequency obtained from modal analysis for this disk is 68.62 MHz (*Mode_3* in Fig.7); the simulation is performed by sweeping the frequency within a range of say, 50 to 100 MHz. The contribution of a few nearby unwanted modes (*Mode_1, 2, 4* and *5*) to the response is also shown in the graph. The peak in the frequency response spectrum corresponds to the desirable frequency of the breathing mode.

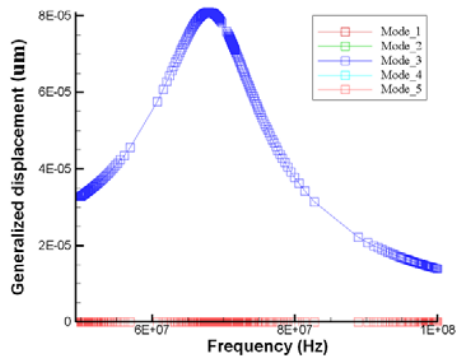

Figure 7. Structural response of the disk resonator subjected to a harmonic excitation

## 4. MODIFIED RESONATOR GEOMETRIES

Another alternative disk resonator structure is shown in Fig. 8. It is a disk resonator devoid of the stem at its center. Instead, the disk is clamped to the anchor by two thin support beams attached to its boundary. This kind of structure is the only option if we need to use the SOI process for fabricating disk resonators with increased thickness of 25 μm. A square extensional (SE) bulk-acoustic mode microresonator has also been designed as shown in the following Fig. 9. In the SE mode, the square plate contracts and extends symmetrically on all its four sides on being excited through lateral capacitive-gap drive electrodes present on each side of the structure, as shown in the figure.

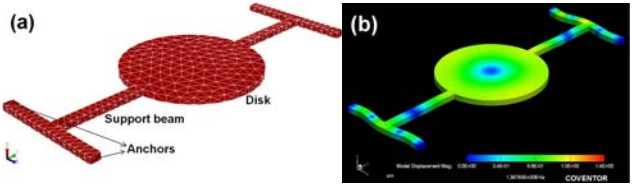

Figure 8. (a) Schematic diagram of a resonator disk supported at the periphery. (b) The fundamental radial-contour modal shape of the same.

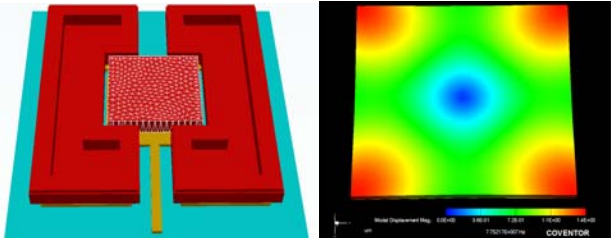

Figure 9 Schematic view of the square-shaped disk resonator. The vibrating portion is highlighted with its meshing. Its square extensional modal shape is shown in the right.

## 5. FABRICATION

The resonators have been fabricated using the PolyMUMPs [6] process. The Thickness of the resonators is 2μm due to process constraint. DC bias connection and RF-feed connections are implemented by the POLY0 layer and the disk structure and side actuating-electrodes are realized with the POLY1 layer. The side electrodes are clamped with the POLY0 layer by creating an ANCHOR and the disk structure is fixed with the help of a stem to the POLY0 layer. Fig.10. shows a field emission scanning electron microscope (FESEM) image of the disk resonator and Fig.11. shows the scanning electron microscope (SEM) images of resonators with modified geometry.

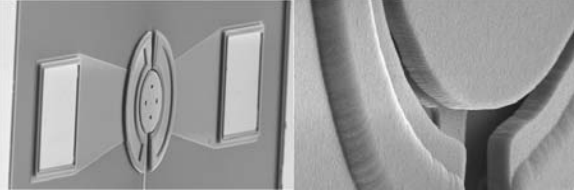

Figure 10 FESEM micrographs of a disk resonator taken at an inclination angle of about 70º

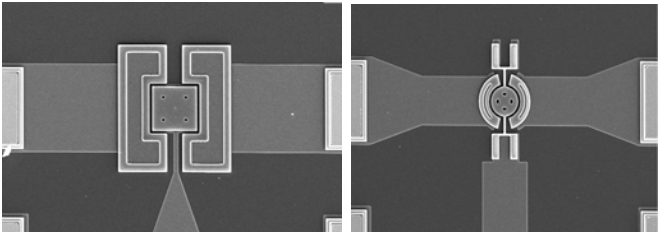

Figure11. SEM micrographs of resonators with modified geometry

# 6. CONCLUSION

The MEMS based circular-disk-resonator for RF wireless applications has been designed and simulated in this work along with detailed discussion of the working principle and theory behind micromechanical disk resonators. A few other bulk mode modified geometries have been designed and simulated also. Mechanical simulation results found from CoventorWare tool have been found to match quite well with the theoretical values. The disk resonators have been tested in laser Doppler vibrometry (LDV) for their primary mechanical responses. Out of plane excitation has been excited in the structure and has been recorded in LDV (Fig. 12.). The results show that these structures were properly released. These mechanical vibrations do not correspond to the desired mode shapes of the disk as a resonator. This is due to the fact that LDV can only measure the out-of-plane vibrations, whereas the desired modes of vibration of the disks are in-plane. So LDV measurements cannot reveal functionality of these structures. Detailed electrical characterization of these structures will be reported soon. The performance of the resonators can certainly be further enhanced (e.g., reduced series motional resistance and higher resonance frequencies) by going for a MEMS fabrication process where lower feature-sizes and spacings (~ 100 nm) can be achieved.

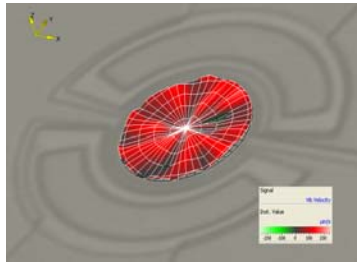

Figure12. Laser Doppler Vibrometry (LDV) of a disk resonator confirming the proper release.

# ACKNOWLEDGEMENTS

The work presented here is supported by National Programme on Micro and Smart Systems (NPMASS), Govt. of India. The authors would like to express their deep gratitude to Prof. Rudra Pratap and Prof. Navakanta Bhat for extending the infrastructure of characterization at Indian Institute of Science,Bangalore.